# Transferable Named Sparse Features for Morphology-based Gleason Grading

Argyrios Zafeiriou[a,*], Athanasios Kallipolitis[a], Ilias Maglogiannis[a]

[a]Department of Digital Systems, University of Piraeus, 80 Karaoli and Dimitriou Street, 18534 Piraeus, Greece

[*]Corresponding author. E-mail address: zafeiriou@unipi.gr (A. Zafeiriou)

## Abstract

Deep learning systems grade prostate biopsies at the level of pathologists, but most report a grade without the glandular architecture that justifies it. This work presents a grading system built on a sparse code over the tokens of a pathology foundation model. A sparse autoencoder pretrained on biopsies is aligned with multi-reader annotations, so that 69 of its 2048 features are named after Gleason morphologies such as cribriform glands, cords, and comedonecrosis. The same code feeds a grading network, which reads all features to predict Gleason patterns and the Grade Group, while morphology maps read directly from the named features. We evaluate the system on five public datasets that include needle biopsies, tissue-microarray cores, multi-reader annotations, and three scanners. The morphology maps remain stable across alignment runs and, without retraining, detect cribriform glands in external biopsies (area under the receiver operating characteristic curve, AUROC, 0.849) and pattern 5 in unseen cores when evaluated by up to six pathologists (AUROC 0.846 to 0.919). In an independent biopsy cohort, Grade Group agreement with three pathologists on Leica scans is close to their agreement with each other. Removing the named features reduces the corresponding Gleason pattern predictions by 49% to 98%, with effects larger than those from 20 activity-matched random feature removals. These results show that a shared sparse representation can support competitive Gleason grading while providing named morphological evidence that transfers across datasets and whose contribution to grading can be measured directly.

# 1 Introduction

Gleason grading describes prostate cancer by its glandular architecture, where tumor regions are assigned Gleason patterns (GPs) 3, 4 and 5. The most extensive and the highest pattern on a biopsy define the Gleason score and its Grade Group (GG) from 1 to 5 (Epstein et al., 2016; European Association of Urology, 2026; van Leenders et al., 2020). Architectural patterns may carry prognostic information beyond the aggregate grade. For example, cribriform glands in pattern 4 are associated with adverse outcomes (Kweldam et al., 2015; van der Kwast et al., 2021), and comedonecrosis in pattern 5 with worse outcomes after surgery (Atallah et al., 2021; Hansum et al., 2021). Additionally, expert disagreement is frequent near adjacent Gleason pattern boundaries, including recognition of cribriform architecture and pattern 5 (Allsbrook et al., 2001; Mittmann et al., 2025; Shah et al., 2015), so reference annotations from several readers carry genuine expert variability. For automated grading systems, therefore, tissue evidence behind a grade matters as much as the grade itself for clinical trust and usability.

In benchmark and widely used datasets like the PANDA challenge, deep learning (DL) algorithms reach pathologist-level agreement in Gleason grading on external validation (Bulten et al., 2022). At the same time, large-scale trained pathology foundation models (FMs) now provide histological representations (Chen et al., 2024; Yan et al., 2025) that enhance downstream tasks like classification and segmentation with out-of-the-box strong encoding capabilities. Both classic DL and FM-based models rely on complex learned computations that obscure how inputs are converted into a diagnosis, while European guidance on trustworthy artificial intelligence and the EU Artificial Intelligence Act emphasize transparency and human oversight for high-risk systems (European Parliament and Council of the European Union, 2024; High Level Expert Group On Artificial Intelligence, 2019).

Most post-hoc image explanations assign importance to pixels or regions (Bach et al., 2015; Chattopadhyay et al., 2018; Samek et al., 2021; Selvaraju et al., 2017). In pathology, such maps may localize relevant tissue without naming the architectural structures used by pathologists (Graziani et al., 2020), which is the objective of concept-based methods. Concept bottleneck models (CBMs) predict predefined concepts and derive the target from them (Koh et al., 2020), and Testing with Concept Activation Vectors (TCAV) measures the sensitivity of a trained model to concept directions without changing it (Kim et al., 2017). In prostate cancer, Mittmann et al. (2025) predict pathologists' histological explanations and derive Gleason patterns from them. Sparse autoencoders (SAEs) offer a representation-level alternative by decomposing dense embeddings into a small number of active features (Cunningham et al., 2023; Pach et al., 2025). Recent pathology studies link sparse features with cell types, tissue components, morphology, and artifacts (Le et al., 2024; Zhao et al., 2026), while mechanistic interventions test model dependence on such features (Patra et al., 2026). It remains unknown whether such sparse and named features can be tied to the named Gleason morphologies and whether a competitive grading model can use them.

In this work, we build a grading and GP recognition system on a sparse code trained over UNI2-h tokens and aligned on multi-reader labels of six named morphologies (well-formed glands, cribriform glands, poorly formed and fused glands, cords, solid tumor groups and comedonecrosis). The grading network reads the full code, the activations of the named features give a map for each morphology, and removing the named features before the grading network measures how much the grade of each pattern depends on them.

We evaluate the system on five public datasets that cover needle biopsies, tissue-microarray cores, image patches, multi-reader annotations, and three scanners. This work makes four contributions:

1. A grading system in which one sparse code feeds both the grading network and the morphology maps, so the evidence for a grade is read from the representation from which the grade is computed.
2. Evidence that the code transfers across cohorts, image types and scanners. It keeps the sparsity it was trained with on every dataset and scanner, its maps transfer to external biopsies and unseen cores without retraining, and five of its six maps keep their accuracy across runs that include independently pretrained SAEs.

3. A grading network that retains the competency of a dense counterpart on held-out biopsies, transfers better to an external hospital and agrees with three pathologists close to their agreement with each other on an independent cohort.
4. A direct test of dependence. Removing the named features of a pattern before the grading network removes 49% to 98% of its predictions, and each named removal ranks first against 20 random sets of unnamed features of equal size and activity.

# 2 Related Work

Automated Gleason grading has progressed from direct pattern prediction to models that expose clinically relevant tissue architecture. Silva-Rodríguez et al. (2020) integrate Gleason grading with cribriform detection by using a network specifically trained to detect the cribriform pattern. More general concept-based approaches connect model predictions to human-interpretable concepts. TCAV evaluates the sensitivity of a trained model to concept directions without altering its parameters (Kim et al., 2017), while concept bottleneck models derive final predictions strictly from intermediate concept bottlenecks (Koh et al., 2020). These works illustrate two approaches to concept-based interpretation. TCAV probes a trained representation, while CBMs route predictions through explicit concept representations.

SAEs provide another way to interpret pathology representations. Le et al. (2024) decompose pathology FM encodings into sparse features associated with biological structure, morphology, and technical variation. Zhao et al. (2026) use a sparse decomposition as a compositional representation of histology and combine it with expert interpretation to produce spatial tissue descriptions. ProtoMIL uses SAE-derived pathology concepts as the representation of an interpretable whole-slide image (WSI) classifier, allowing users to intervene on those concepts (Sun et al., 2026). These studies demonstrate how sparse features can reveal meaningful histology and contribute to downstream prediction, while further literature has studied the role of sparse spaces as an intervention. Patra et al. (2026) combine SAEs with activation patching to identify mechanisms associated with adversarial vulnerability in digital pathology. In prostate cancer, EXPOSE identifies sparse dimensions associated with acquisition domain and suppresses them prior to relapse prediction to improve robustness across acquisition artifacts (Witte et al., 2026).

The semantics of an SAE should also not be assumed to be tied to a unique feature dictionary, as Paulo and Belrose (2025) show that SAEs trained on the same activations with different random seeds can recover different feature sets, while Leask et al. (2025) show more generally that SAE latents need not form a complete or canonical set of units. Work on monosemanticity showed that SAEs can recover interpretable features from transformer activations, both for small- and large-scale modeling (Bricken et al., 2023; Templeton et al., 2026). Interventions on these features can alter model behavior in ways consistent with their interpretations, supporting their use as meaningful units of analysis without implying that the recovered feature dictionary is unique or complete.

In this work, we use the GleasonXAI (Mittmann et al., 2025) dataset, which provides expert annotations of prostate cancer morphologies on 1015 tissue-microarray (TMA) cores. Mittmann et al. (2025) train an interpretable CBM-like dense segmentation network against soft labels of the experts' morphology annotations, which account for disagreement between readers, and evaluate it on a randomized train-validation-test split of the cores. Instead, we first fit a generic prostate cancer SAE on the large-scale WSI PANDA dataset (Bulten et al., 2022), align it to TMA annotations that identify and name morphological features, and evaluate morphology localization directly from the activations of the aligned sparse features. The complete sparse representation remains available to the Gleason segmentation network, while morphology maps are computed directly from the aligned feature sets. This separates concept localization from unrestricted grading, in contrast with CBM architectures, where grading stems from concept predictions. The concept-grade relation is quantified by intervening during inference, ablating the named features and inspecting the predicted grade. Additionally, we include random feature interventions matched on activation frequency to distinguish feature importance from catastrophic perturbation. For us, intervention upon SAE features allows us to establish their causal contribution to the model's predictions, while we assess the stability of the resulting morphology maps across independent SAE pretraining and concept alignment runs. The robustness and limitations of the system are evaluated using external cohorts and a range of imaging types, including TMA cores and WSIs from various scanners and scales.

# 3 Methods

## 3.1 System overview

Figure 1 summarizes the architecture. A sparse autoencoder turns UNI2-h tokens into 2048 sparse features, which we call the code, and alignment with GleasonXAI names 69 of them in six morphology sets. A U-Net ensemble, the grading network, reads the full code to give the Gleason pattern map and the GG. The six morphology maps are read from the named sets, and removing named features before the grading network shows how much the grade depends on them.

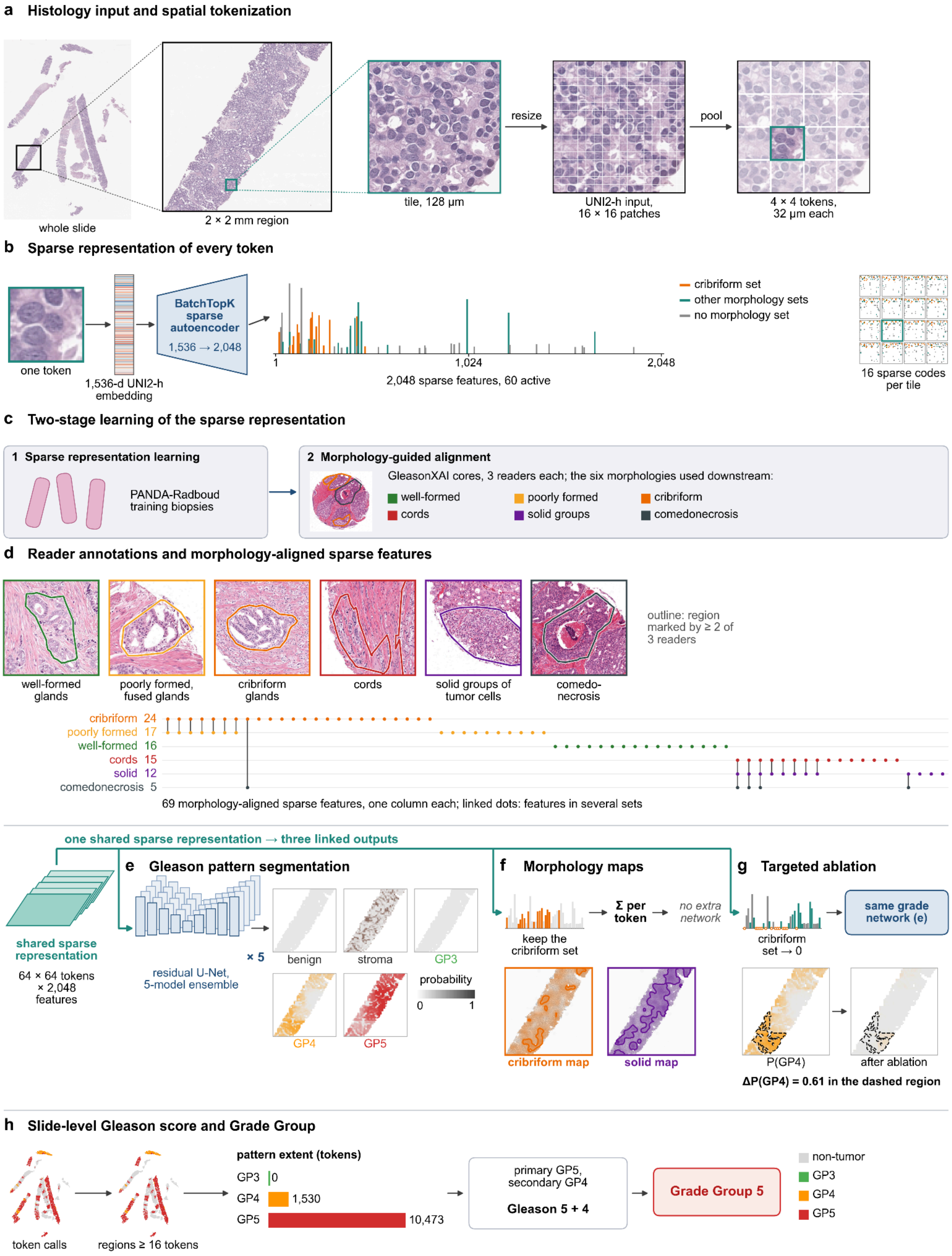

*Figure 1: Architecture of the system. (a) Tissue is tiled into 128 µm tiles, encoded by UNI2-h and pooled into 4 × 4 tokens of 32 µm. (b) A BatchTopK SAE maps each 1536-dimensional token to 2048 sparse features, 60 of which are active in this example token. (c) The SAE is pretrained on PANDA-Radboud biopsies and aligned with GleasonXAI annotations of three readers per core. (d) Reader outlines of the six morphologies used downstream and their 69 named features, with linked dots for features shared between sets. (e) The grading network, an ensemble of five residual U-Nets, reads the full code and gives the probability of each class. (f) A morphology map is the sum of its named features at each token, with no extra network. (g) Removing the cribriform set before the*

*grading network lowers the pattern 4 probability by 0.61 in the outlined region. (h) The slide-level rule, from token calls to regions of at least 16 tokens, pattern extents, the Gleason score and the GG.*

## 3.2 Datasets

We use five public datasets of hematoxylin and eosin (H&E) stained prostate tissue (Table 1, Figure 2). PANDA-Radboud (Bulten et al., 2022) provides needle-biopsy WSIs with semi-automatic pixel masks and public slide GG labels, and is used to pretrain the SAE and to train and select the grading networks. GleasonXAI (Mittmann et al., 2025) provides TMA cores on which panels of three pathologists, referred to as readers, outlined nine morphology classes, and is used to align the code. Gleason2019 (Nir et al., 2018) provides TMA cores graded by three to six pathologists each. Of its 244 cores, 143 are also part of GleasonXAI, and the other 101, the unseen cores, were never used for alignment, training, or model selection. SICAPv2 (Silva-Rodríguez et al., 2020) provides biopsy patches from an external hospital with pattern masks and cribriform labels. PAR (Muhammad Ali et al., 2026) provides 339 biopsies from 185 patients, each scanned on a Leica, a Hamamatsu, and a Grundium scanner and graded on screen by up to three pathologists, referred to as raters. Gleason2019, SICAPv2, and PAR are used only for evaluation.

*Table 1: Public datasets used in this study.*

| Dataset | Material | Size | Annotation | Use |
|---|---|---|---|---|
| PANDA-Radboud | Needle-biopsy WSIs | 5060 slides (4048 train, 506 validation, 506 test) | Semi-automatic masks for stroma, benign epithelium and patterns 3 to 5. Slide GG | SAE pretraining, grading network training and selection, test |
| GleasonXAI | TMA cores from three sources | 1015 cores (812 alignment, 203 held out) | Nine morphology classes outlined by three readers per core | Alignment, morphology evaluation |
| Gleason2019 | TMA cores | 244 cores (143 in GleasonXAI, 101 unseen) | Gleason pattern maps from three to six pathologists per core | Pattern 5 evaluation |
| SICAPv2 | Biopsy patches | 155 slides, 95 patients. Test partition: 2122 patches, 31 slides, 21 patients | Pattern masks, patch labels, cribriform labels for pattern 4 patches | External grading, cribriform detection, feature removal |
| PAR | Needle-biopsy WSIs from three scanners | 339 slides, 185 patients. 323 slides scored on all three scanners | Slide Gleason score and GG from up to three raters | GG agreement, scanner comparison |

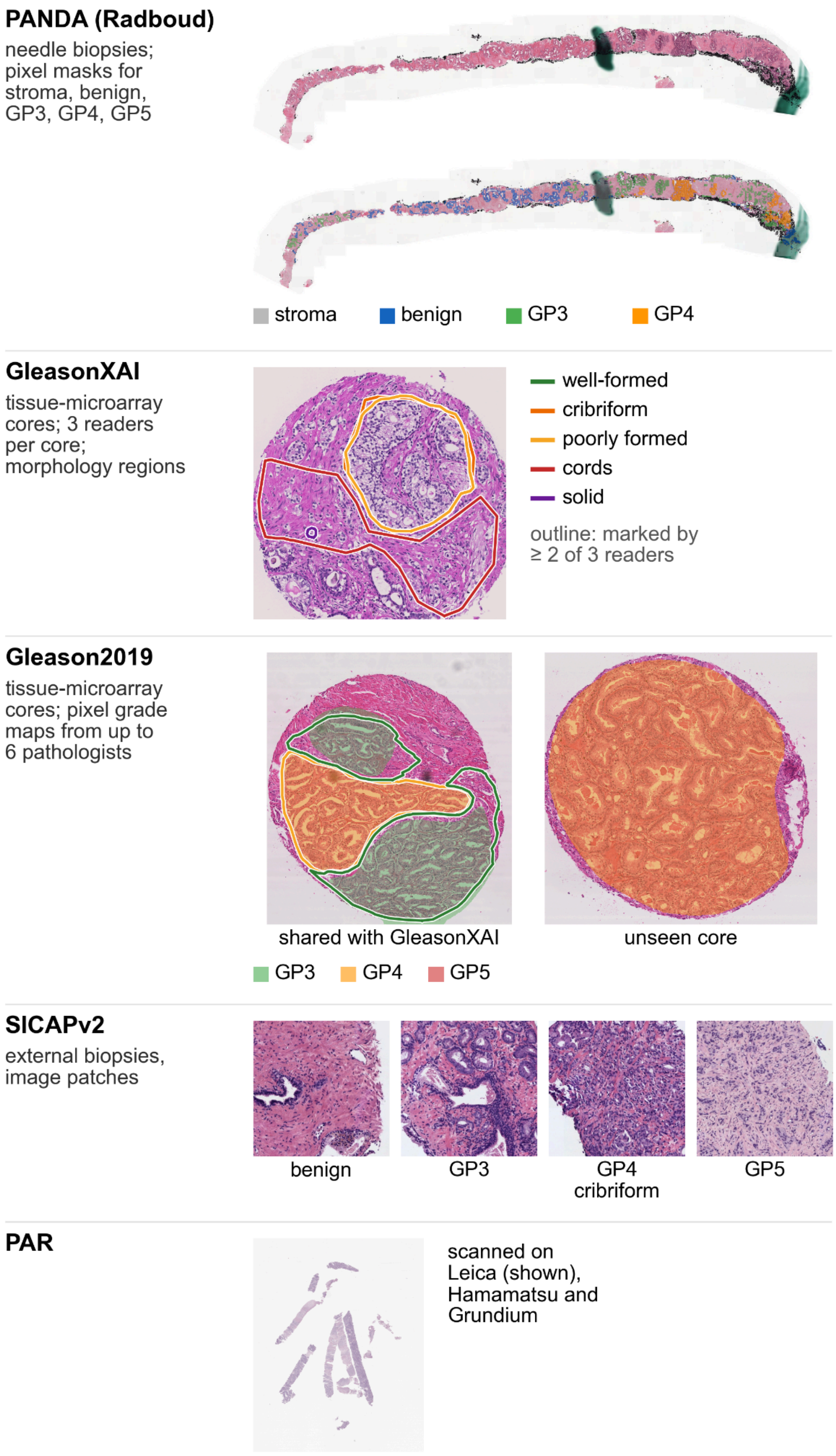

*Figure 2: Datasets. PANDA-Radboud trains and selects the SAE and the grading network, GleasonXAI aligns the code, and Gleason2019, SICAPv2 and PAR are used only for evaluation.*

## 3.3 Sparse code

All images are tiled into regions of approximately 128 µm, resized to 224×224 pixels, and encoded by UNI2-h (Chen et al., 2024; Mahmood Lab, 2025) with its standard input normalization. No stain normalization is applied. From the 18th of the encoder's 24 transformer blocks, we take the 16×16 patch tokens of each tile and average them in non-overlapping 4×4 groups. Each tile thus gives 16 tokens of 1536 dimensions, each covering about 32 µm of tissue at a fixed position (Figure 1a). Mostly empty tiles of WSIs are discarded, and on cores and SICAPv2 patches, tissue is estimated per token.

A Matryoshka BatchTopK SAE (Bussmann et al., 2024, 2025) maps each token to 2048 non-negative features and reconstructs it from a dictionary of unit-norm feature directions (Figure 1b-c). In training, it keeps the 48×B largest activations of a batch of B tokens, so a token has 48 active features on average. At inference, a fixed activation threshold set to the same average activity replaces this batch selection. The SAE is pretrained on the 4048 PANDA-Radboud training slides to reconstruct their UNI2-h tokens, with an auxiliary loss that keeps features from becoming permanently inactive. We select it by the fraction of variance unexplained on the 506 validation slides.

The pretrained SAE is then aligned on 812 GleasonXAI cores (Figure 1c). For each of the nine morphology classes, a gated linear layer on the code predicts, at each tissue token, the fraction of the three readers who marked that class. The encoder, the decoder, and the gated layers are trained together. Reconstruction and a rehearsal loss on PANDA-Radboud tokens keep the code fitted to biopsies, and a sparsity penalty on the gates limits the number of features each class can use. A feature joins the set of a class when its gate for that class ends above 0.5. Table 2 lists the losses and training settings of the SAE and of its alignment.

Three of the nine classes (compressed or angular glands, glomeruloid glands and single cells) were marked by at least two readers on too few tokens to be modeled reliably, at most 1362 each against at least 3956 for the others. They still supervise the alignment, and the other six are the morphologies used downstream (Figure 1c). Their sets contain 24 cribriform, 17 poorly formed and fused, 16 well-formed, 15 cord, 12 solid-group and 5 comedonecrosis features (Figure 1d). Sets may share features, and comedonecrosis and solid groups share most of their features with cords, as these morphologies co-occur in the readers' drawings. The union of the six sets holds 69 named features, and the other 1979 features are unnamed. By pattern, the named features form a pattern 3 set of 16 (well-formed glands), a pattern 4 set of 34 (cribriform and poorly formed glands) and a pattern 5 set of 20 (cords, solid groups and comedonecrosis).

*Table 2: Losses and training settings of the sparse code. FVU: fraction of variance unexplained.*

| Setting | Value |
|---|---|
| SAE type and loss | Matryoshka (Bussmann et al., 2025) BatchTopK (Bussmann et al., 2024) SAE. Mean squared error averaged over the first 512, the first 1024 and all 2048 features, plus the auxiliary loss of Bussmann et al. (2024) with weight 1/32 |
| SAE input | Tokens of the 18th block before the final layer normalization, centered and scaled to a root-mean-square norm of $\sqrt{1536}$ with statistics of the training slides |
| Pretraining | Adam, learning rate $5 \times 10^{-4}$ (constant), gradient clipping at 1.0, 4096 tokens per step, 30 epochs of 19.1 million tokens. Epoch 29 selected (validation FVU 0.173) |
| Inference threshold | Set so that 48 features are active on average over 262144 tokens of the training slides. 2.085 for the aligned code |
| Alignment loss | FVU on core tokens + 1.0 × FVU on rehearsal tokens + 4.0 × binary cross-entropy between predicted and observed vote fraction + 3.0 × expected number of open gates / (9 × 2048) |
| Gates | Hard-concrete (Louizos et al., 2017) with $\beta = 2/3$, $\gamma = -0.1$, $\zeta = 1.1$ and an initial open probability of 0.9 |
| Alignment target | A reader marks a token when the outline covers at least half of it. Tissue tokens only (tissue fraction ≥ 0.4) |
| Alignment training | AdamW, cosine schedule, gradient clipping at 1.0, 25 epochs, final epoch used. 4096 core tokens and 4096 rehearsal tokens per step. Learning rate $3 \times 10^{-4}$ and weight decay $10^{-4}$ for the SAE and the gated linear layers. Learning rate $5 \times 10^{-2}$ without weight decay for the gates |
| Rehearsal pool | 640000 tokens from the PANDA-Radboud training slides |
| Feature scale of the maps | Root-mean-square activation of each feature on 32768 tokens of the alignment cores |
| Map scale | 90th percentile of the summed map over the tissue tokens of the training cores of the same alignment |

## 3.4 Grade and morphology maps

The grading network is an ensemble of five residual U-Nets that read the full code and predict, at each token, independent probabilities for benign epithelium, stroma and patterns 3, 4 and 5. Each network is trained on four folds of a five-fold split of the PANDA-Radboud training slides and selected on the fifth, against token labels from the masks, with binary cross-entropy, soft Dice loss and deep supervision. Inference runs on windows of 64×64 tokens with a stride of 32, and the probabilities are averaged over overlapping windows and over the five networks. Class thresholds are chosen on the validation slides. A token is called for every class whose probability exceeds its threshold. For display and for the GG, each token also receives a single label, the called pattern with the largest probability relative to its threshold, or non-tumor when no pattern is called.

The GG follows from the single labels by a fixed rule (Figure 1h). Tokens with the same pattern are grouped into 8-connected regions, and regions smaller than 16 tokens, a size chosen on the validation slides, are discarded. The extent of a pattern is the number of its tokens in the remaining regions. The primary pattern is the pattern with the largest extent, and the secondary pattern is the highest other pattern with a remaining region, or the primary itself when there is none. The pair gives the Gleason score, which the table of the International Society of Urological Pathology (ISUP) maps to the GG. A slide without any remaining region is GG 0. In the biopsy of Figure 1h, pattern 5 covers 10473 tokens, pattern 4 covers 1530 and pattern 3 covers none, so the slide receives Gleason 5 + 4 and GG 5.

In parallel with Gleason segmentation, morphology maps are computed directly from the aligned feature sets without an additional prediction network. For each morphology, feature activations are scaled by fixed feature-specific activation statistics and summed at each tissue token. The resulting map is normalized by a fixed scale estimated from the alignment data (Table 2). Raw maps are used for detection, and one-token Gaussian-smoothed maps are used for GleasonXAI evaluation and display. For pattern 5 analyses, the cords, solid-group, and comedonecrosis maps are combined by taking their maximum at each token.

## 3.5 Feature removal

We measure how much the grade depends on named features by removing them, a form of ablation. In targeted feature ablation, all activations belonging to a selected feature set are set to zero before the representation enters the Gleason grading network. All remaining sparse features and network parameters are unchanged. We remove each of the six sets and the pattern 4 and pattern 5 sets. At the regional level, dependence is measured as the reduction in the relevant Gleason probability after feature removal (Figure 1g). At cohort level, the effect is summarized by the relative change in the number of thresholded pattern calls. As a control for each pattern, we draw 20 sets of unnamed features of the same size and matched activation frequency and remove them in the same way. Named and random removals are compared on a random subset of 100 test slides, common to all removals. Two descriptive shares complement the removal. The activation share is the fraction of a token's total activation that falls on the named features. The input share is the same fraction measured at the input of the grading network, with each feature's activation multiplied by the norm of the first-layer weights that read it. Both shares describe how much of the input the named features make up, while the removal tests whether the grade depends on them.

## 3.6 Evaluation design

PANDA-Radboud training slides train the SAE and the grading network, and its validation slides select the SAE, the class thresholds and the minimum region size. All PANDA-Radboud results are reported on the test slides, and SICAPv2 and PAR are evaluated with these settings unchanged.

Morphology maps are evaluated on GleasonXAI without reusing any reader panel. Five folds keep all cores of a panel together, so every evaluable core receives its map from an alignment that never saw its panel. All other analyses use the primary alignment on 812 cores, and the 203 held-out cores serve the lift and code-fit analyses. The 143 Gleason2019 cores that are part of GleasonXAI are scored with an alignment trained without them.

Five reference models serve as comparators. The embedding U-Net has the architecture of the grading network but reads the pooled UNI2-h tokens directly, with its own thresholds and the same GG rule. The single sparse U-Net has the architecture and training recipe of an ensemble member and is trained on all 4048 training slides. These two form a matched pair in which only the input differs. For the morphology maps, the dense morphology U-Net is trained on UNI2-h tokens with the GleasonXAI labels, the six-map U-Net on the six morphology maps, and the dense morphology classifier is a logistic model on UNI2-h tokens trained with the alignment labels.

Stability is tested at two levels, since SAEs trained on the same data with different seeds are known to learn different features (Paulo and Belrose, 2025). The alignment is repeated four times from the same pretrained SAE with different initializations, and three more times on SAEs pretrained independently, and we report the range of each map's panel-disjoint area under the receiver operating characteristic curve (AUROC) over the seven runs. A stable range across independently pretrained SAEs shows that the named morphologies do not depend on one particular dictionary.

On PAR, the agreement of the GG with each rater is measured by quadratic weighted Cohen's κ and set beside the κ between raters on the same slides. For each pair of raters, the grading network's κ with one rater is compared with the κ between the two raters on the slides both graded, which gives six paired comparisons. Scanners are compared on the 323 slides scored on all three instruments. On the other 16, tiling of the Hamamatsu scan gave fragments of at most four tiles, below the minimum of 16 tiles and 32 tissue tokens that a fragment needs to be scored. On each scanner, slide-level detection of cancer, pattern 4 and pattern 5 is scored by the share of tissue called and by the mean probability, and evaluated by AUROC against each rater and the raters' majority. Of these 24 comparisons per scanner, we count those in which the Grundium or Hamamatsu AUROC lies within 0.03 of the Leica value.

### 3.7 Metrics and statistics

Gleason segmentation is scored by the Dice coefficient per class, pooled over all tokens of a cohort and computed on the per-class calls. Macro Dice averages benign epithelium and patterns 3, 4, and 5. Morphology maps and pattern 5 detection are scored by AUROC at the token level, against tokens marked by at least two of the three readers on GleasonXAI and against each pathologist's map on Gleason2019. Cribriform detection on SICAPv2 is scored per patch by the share of the patch that the map calls cribriform. GG agreement is measured by quadratic weighted κ over GG 0 to 5, exact agreement, and agreement within one group.

The lift of a feature for a morphology is the readers' vote fraction for that morphology averaged over tokens with the feature's activation as weight, divided by its unweighted average over the same tokens. It is computed on the 203 held-out cores. A set's lift uses the sum of its features, and a feature's highest-lift morphology is the one with the largest lift. Code fit is described by the mean number of active features per token, the median relative reconstruction error, and the median share of activation carried by the named features. The relative reconstruction error is the squared norm of the reconstruction error divided by the squared norm of the token. Shares are reported on tumor tokens, those called pattern 3, 4 or 5 by the grading network, and on the remaining tissue.

Intervals are 95% percentile bootstrap intervals from 2000 replicates, or 1000 for the computationally intensive analyses. We resample slides on PANDA-Radboud and SICAPv2, cores on GleasonXAI and Gleason2019, and patients on PAR. Paired comparisons use the same replicates for both models.

# 4 Results

Tables 3 to 9 summarize the quantitative results, and Figures 3 to 6 showcase examples from each analysis.

### 4.1 Fit of the sparse code

Table 3 reports how well the code fits each cohort. The code keeps its sparsity on every imaging type, although it was pretrained only on PANDA-Radboud biopsies. The mean number of active features per token ranges from 46.7 to 51.9, close to the 48 targeted in training, and the median relative reconstruction error ranges from 0.16 to 0.27, on biopsies, TMA cores, SICAPv2 patches and the three PAR scanners alike. The lowest error, 0.16, is observed on the unseen Gleason2019 cores. On tumor tokens, the named features carry a median share of the activation from 0.38 to 0.45 on every cohort apart from the Hamamatsu scans, where the share is 0.31, while on the remaining tissue it ranges from 0.00 to 0.17. The 69 named features thus carry a large part of the code wherever the tumor is called, and the Hamamatsu scanner is the one condition under which this share drops.

*Table 3: Fit of the code per cohort.*

| Cohort | Active features per token, mean | Relative reconstruction error, median | Activation share, tumor | Activation share, other tissue |
|---|---|---|---|---|
| PANDA-Radboud test | 48.1 | 0.21 | 0.42 | 0.12 |
| SICAPv2 test | 50.4 | 0.25 | 0.41 | 0.17 |
| PAR Leica | 46.7 | 0.26 | 0.38 | 0.04 |
| PAR Grundium | 50.8 | 0.27 | 0.38 | 0.06 |
| PAR Hamamatsu | 48.7 | 0.25 | 0.31 | 0.03 |
| GleasonXAI alignment cores | 51.9 | 0.19 | 0.45 | 0.04 |
| GleasonXAI held-out cores | 51.6 | 0.19 | 0.45 | 0.03 |
| Gleason2019 unseen cores | 47.0 | 0.16 | 0.42 | 0.00 |

## 4.2 Gleason grading

On the PANDA-Radboud test set, the grading network reaches a macro Dice of 0.730 [0.710, 0.748], with class values between 0.617 for pattern 5 and 0.794 for pattern 4 (Table 4). In the matched comparison, the single sparse U-Net reaches 0.728 [0.708, 0.746] and the embedding U-Net 0.742 [0.722, 0.759], which gives a paired difference of −0.014 [−0.018, −0.009]. The picture changes on the SICAPv2 test slides, where all models use the thresholds chosen on PANDA-Radboud. The grading network reaches Dice values of 0.524, 0.658 and 0.535 for patterns 3, 4 and 5, and the single sparse U-Net exceeds the embedding U-Net by 0.035 [0.007, 0.079] for pattern 4 and by 0.068 [0.015, 0.119] for pattern 5. This means that the cost of the code on PANDA-Radboud turns into a gain for patterns 4 and 5 on the biopsies of an external dataset, indicating a more robust representation that transfers better to an external hospital.

*Table 4: Token-level Gleason segmentation, Dice.*

| Cohort | Class | Grading network | Single sparse U-Net | Embedding U-Net |
|---|---|---|---|---|
| PANDA-Radboud | Benign | 0.769 [0.752, 0.784] | 0.766 [0.749, 0.781] | 0.784 [0.769, 0.799] |
| | Stroma | 0.962 [0.959, 0.965] | 0.961 [0.958, 0.964] | 0.965 [0.962, 0.968] |
| | Pattern 3 | 0.742 [0.714, 0.766] | 0.746 [0.719, 0.770] | 0.749 [0.722, 0.772] |
| | Pattern 4 | 0.794 [0.775, 0.811] | 0.791 [0.772, 0.809] | 0.803 [0.785, 0.819] |
| | Pattern 5 | 0.617 [0.549, 0.675] | 0.610 [0.543, 0.667] | 0.632 [0.565, 0.689] |
| | Macro | 0.730 [0.710, 0.748] | 0.728 [0.708, 0.746] | 0.742 [0.722, 0.759] |
| SICAPv2 | Pattern 3 | 0.524 [0.400, 0.598] | 0.499 [0.351, 0.578] | 0.472 [0.335, 0.551] |
| | Pattern 4 | 0.658 [0.565, 0.712] | 0.658 [0.569, 0.714] | 0.623 [0.509, 0.691] |
| | Pattern 5 | 0.535 [0.285, 0.634] | 0.507 [0.285, 0.614] | 0.439 [0.222, 0.551] |

At slide level, the GG derived from the grading network agrees with the PANDA-Radboud labels at κ 0.784 [0.732, 0.830], with an exact agreement of 0.577 [0.534, 0.621] and an agreement within one group of 0.881 [0.852, 0.907] (Table 5). The matched comparison favors the code at this level, with κ 0.803 [0.751, 0.847] for the single sparse U-Net against 0.762 [0.710, 0.812] for the embedding U-Net, a paired difference of 0.041 [0.008, 0.076]. Note that the single sparse U-Net agrees slightly better with the labels than the ensemble. The ensemble was adopted before any test result, on its higher validation Dice.

*Table 5: Slide-level GG agreement.*

| Cohort | Comparison | Measure | Slides (patients) | Value [95% CI] |
|---|---|---|---|---|
| PANDA-Radboud | Grading network vs label | κ | 506 | 0.784 [0.732, 0.830] |
| | Grading network vs label | Exact agreement | 506 | 0.577 [0.534, 0.621] |
| | Grading network vs label | Within one group | 506 | 0.881 [0.852, 0.907] |
| | Single sparse U-Net vs label | κ | 506 | 0.803 [0.751, 0.847] |
| | Embedding U-Net vs label | κ | 506 | 0.762 [0.710, 0.812] |
| | Single sparse − embedding | Paired Δκ | 506 | 0.041 [0.008, 0.076] |
| PAR Leica | Grading network vs rater 1 | κ | 339 (185) | 0.804 [0.748, 0.852] |
| | Grading network vs rater 2 | κ | 337 (184) | 0.746 [0.684, 0.801] |
| | Grading network vs rater 3 | κ | 59 (53) | 0.795 [0.667, 0.887] |
| | Rater 1 vs rater 2 | κ | 337 (184) | 0.811 [0.763, 0.854] |
| | Rater 1 vs rater 3 | κ | 59 (53) | 0.836 [0.696, 0.928] |
| | Rater 2 vs rater 3 | κ | 58 (52) | 0.751 [0.606, 0.862] |
| PAR Grundium | Grading network vs rater 1 | κ | 323 (183) | 0.812 [0.754, 0.862] |
| | Grading network vs rater 2 | κ | 321 (182) | 0.795 [0.730, 0.848] |
| | Grading network vs rater 3 | κ | 56 (51) | 0.734 [0.557, 0.877] |
| | GG against Leica GG | κ | 323 (183) | 0.872 [0.815, 0.920] |
| PAR Hamamatsu | Grading network vs rater 1 | κ | 323 (183) | 0.660 [0.574, 0.734] |
| | Grading network vs rater 2 | κ | 321 (182) | 0.634 [0.545, 0.714] |
| | Grading network vs rater 3 | κ | 56 (51) | 0.729 [0.545, 0.868] |
| | GG against Leica GG | κ | 323 (183) | 0.741 [0.653, 0.816] |

On the Leica scans of PAR, the GG agrees with raters 1, 2 and 3 at κ 0.804, 0.746 and 0.795, while the raters agree with each other at κ between 0.751 and 0.836. In five of the six paired comparisons, the agreement of the grading network with a rater does not differ detectably from the agreement between the two raters. The exception is rater 2, with whom the grading network agrees 0.065 [0.014, 0.119] less than rater 1 does. On the 323 slides scored on all three instruments, the GG from the Grundium scans agrees with the GG from the Leica scans at κ 0.872 [0.815, 0.920], and the GG from the Hamamatsu scans at 0.741 [0.653, 0.816]. Agreement with the raters follows the same order, between 0.734 and 0.812 on Grundium and down to 0.660 and 0.634 for raters 1 and 2 on Hamamatsu. Slide-level detection of cancer, pattern 4 and pattern 5 gives the same picture. Of the 24 comparisons per scanner, 22 stay within 0.03 of the Leica AUROC on Grundium and 7 on Hamamatsu for the grading network, against 18 and 4 for the embedding U-Net. Both models lose accuracy on the Hamamatsu scans, with the grading network in fewer comparisons than the embedding U-Net.

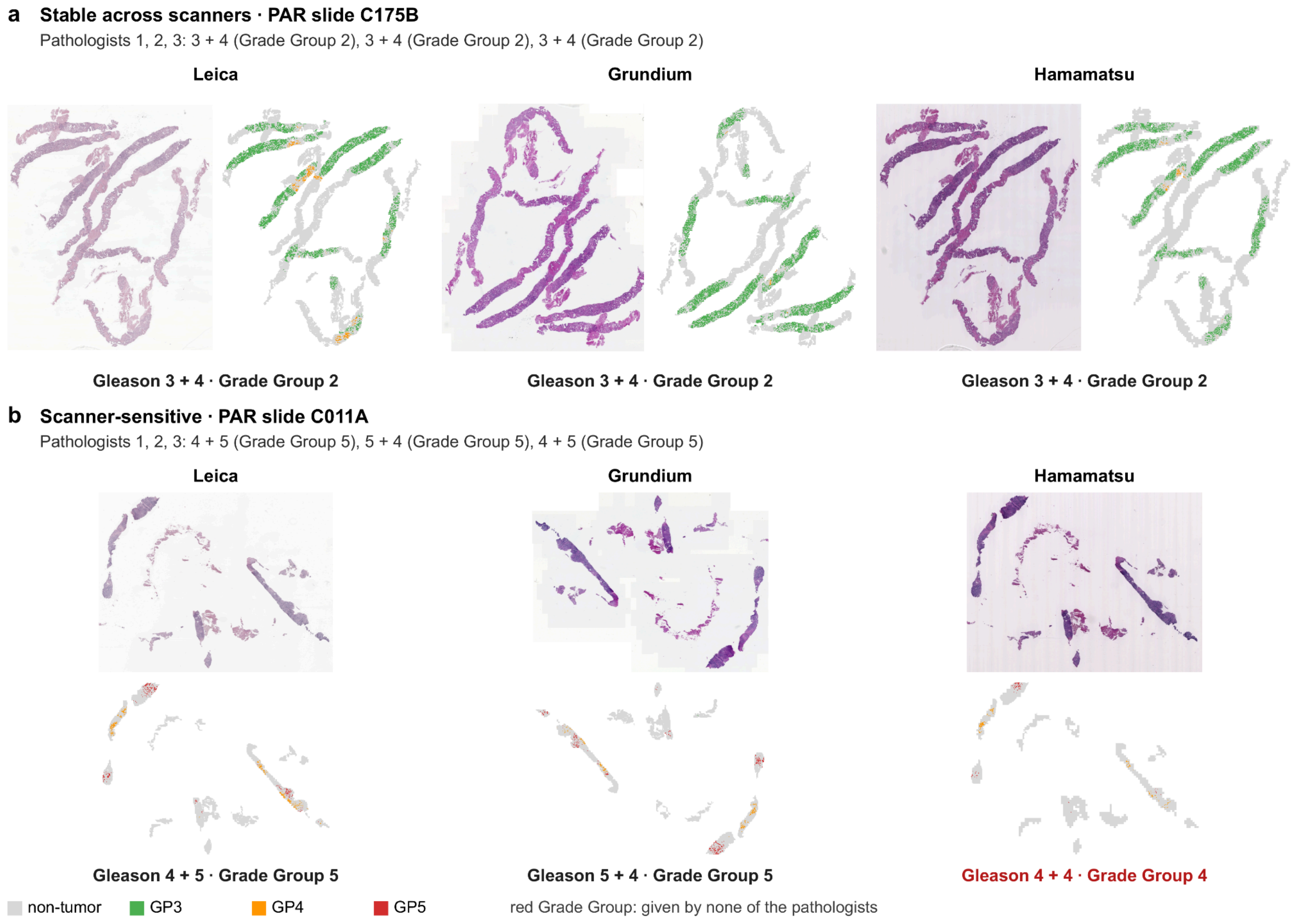

*Figure 3: Grading across scanners. Matched PAR biopsies scanned on the Leica, Grundium and Hamamatsu instruments, with their Gleason pattern maps and GG beside the GG of the three raters. The first example is graded consistently on the three scanners and the second example changes with the scanner.*

## 4.3 Morphology maps

Table 6 summarizes the six morphology maps. On the GleasonXAI cores, with each map taken from an alignment that excluded the panel of the core, the token-level AUROC is 0.949 for cords, 0.943 for solid groups, 0.921 for well-formed glands and 0.899 for cribriform glands, while poorly formed and fused glands reach 0.759 and comedonecrosis 0.719. Across the seven alignment runs, the AUROC of five maps changes by less than 0.01, while that of comedonecrosis ranges from 0.719 to 0.860.

Regarding the specificity of the named feature sets, on the 203 held-out cores, the lift of each set on its own morphology lies between 2.0 for poorly formed and fused glands and 5.0 for cords. At the level of single features, all 16 well-formed features, 14 of the 15 cord features and 20 of the 24 cribriform features have their own morphology as the highest-lift morphology. The solid-group and comedonecrosis sets are less specific, with 3 of 12 and 1 of 5 such features, as they share most of their features with cords.

The cribriform map also transfers to SICAPv2, whose cribriform labels were never used by any part of the system (Table 7, Figure 4). Scored per patch, the map reaches an AUROC of 0.849 [0.723, 0.984] on all test patches and 0.789 [0.629, 0.963] on the pattern 4 patches, where cribriform glands have to be separated from other pattern 4 morphology. The dense morphology U-Net, trained on UNI2-h tokens with the GleasonXAI labels, reaches 0.822 and 0.746, respectively, and the six-map U-Net reaches 0.691 and 0.707. In other words, a sum of named features performs at the level of a network trained with the same labels, on images from a hospital that neither of them has seen.

*Table 6: Morphology detection, stability and specificity.*

| Morphology | Features | Panel-disjoint AUROC [95% CI] | Range over seven runs | Lift on own morphology | Features with own morphology the |
|---|---|---|---|---|---|

| | | | | | highest |
|---|---|---|---|---|---|
| Well-formed glands | 16 | 0.921 [0.908, 0.933] | 0.919 to 0.924 | 3.5 | 16 of 16 |
| Cribriform glands | 24 | 0.899 [0.878, 0.919] | 0.898 to 0.900 | 3.2 | 20 of 24 |
| Poorly formed and fused glands | 17 | 0.759 [0.735, 0.784] | 0.759 to 0.763 | 2.0 | 10 of 17 |
| Cords | 15 | 0.949 [0.929, 0.964] | 0.947 to 0.950 | 5.0 | 14 of 15 |
| Solid groups | 12 | 0.943 [0.924, 0.961] | 0.938 to 0.946 | 4.1 | 3 of 12 |
| Comedonecrosis | 5 | 0.719 [0.617, 0.812] | 0.719 to 0.860 | 4.6 | 1 of 5 |

*Table 7: Cribriform detection on SICAPv2, patch-level AUROC.*

| Patches | Cribriform map | Dense morphology U-Net | Six-map U-Net |
|---|---|---|---|
| All test patches | 0.849 [0.723, 0.984] | 0.822 [0.731, 0.923] | 0.691 [0.527, 0.846] |
| Pattern 4 test patches | 0.789 [0.629, 0.963] | 0.746 [0.642, 0.860] | 0.707 [0.539, 0.852] |

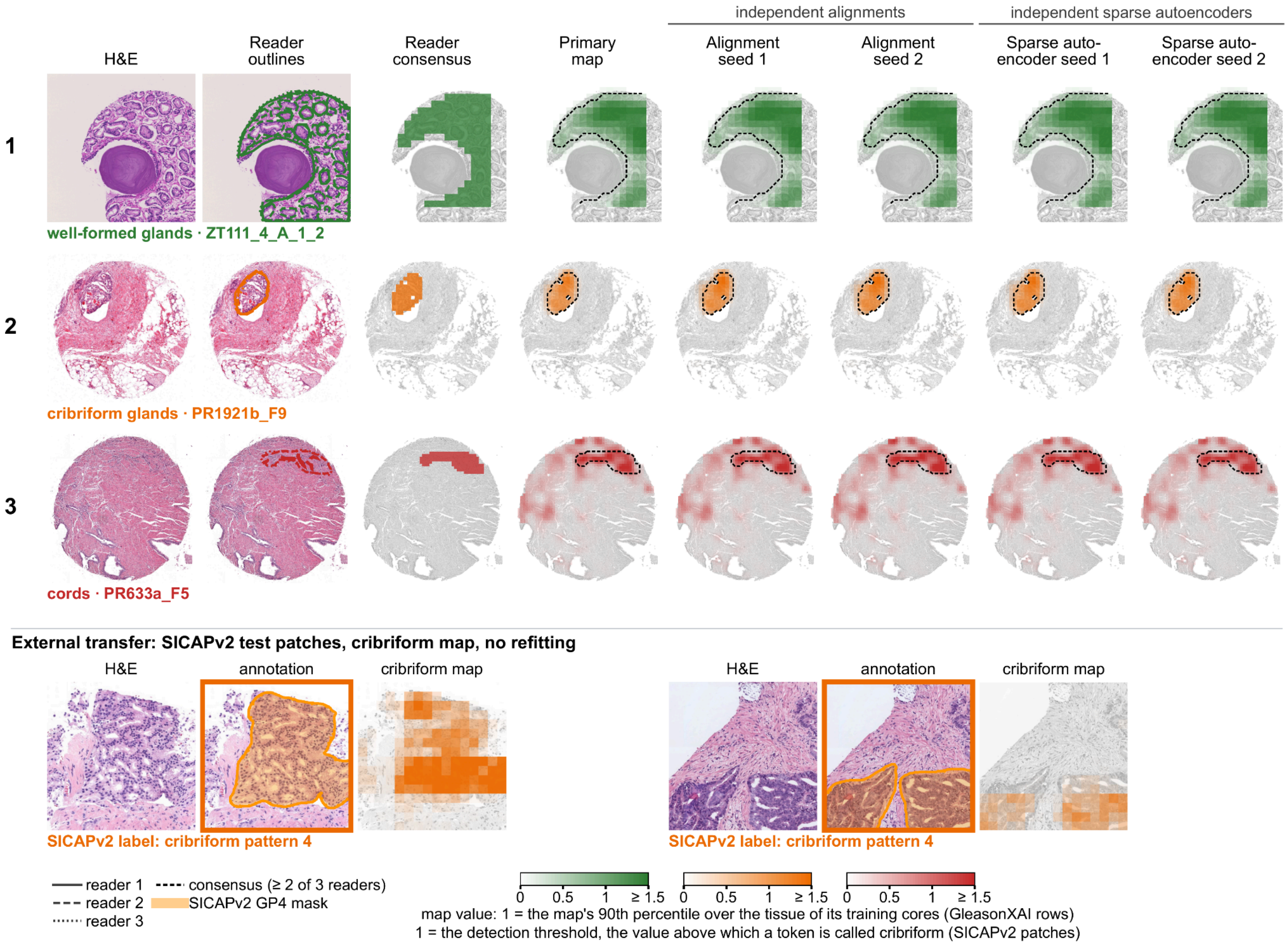

*Figure 4: Morphology maps across training runs and datasets. Maps of representative GleasonXAI cores from repeated alignment and SAE training runs, and the cribriform map on SICAPv2 biopsies.*

## 4.4 Pattern 5 on tissue-microarray cores

The pattern 5 map transfers to TMA cores without retraining (Table 8, Figure 5). On the 101 unseen Gleason2019 cores, it reaches token-level AUROCs from 0.846 to 0.919 against the six pathologists, above the grading network for five of them, and within a narrower range than the dense classifier (0.540 to 0.968).

On the 143 cores shared with GleasonXAI, scored with an alignment that excluded them, it reaches 0.915 to 0.980 against the five pathologists who marked pattern 5, the level of a dense classifier trained with the same labels (0.917 to 0.969). The code thus carries the pattern 5 evidence of the readers' labels without a separate model.

*Table 8: Pattern 5 on Gleason2019 cores, token-level AUROC.*

| Cores | Pathologists | Pattern 5 map | Grading network | Dense morphology classifier | Map - grading network |
|---|---|---|---|---|---|
| 143 cores in GleasonXAI | 5 who marked pattern 5 | 0.915 to 0.980 | 0.599 to 0.689 | 0.917 to 0.969 | 0.282 to 0.349, all 5 intervals above 0 |
| 101 unseen cores | 6 | 0.846 to 0.919 | 0.670 to 0.896 | 0.540 to 0.968 | 0.021 to 0.203, 5 of 6 intervals above 0 |

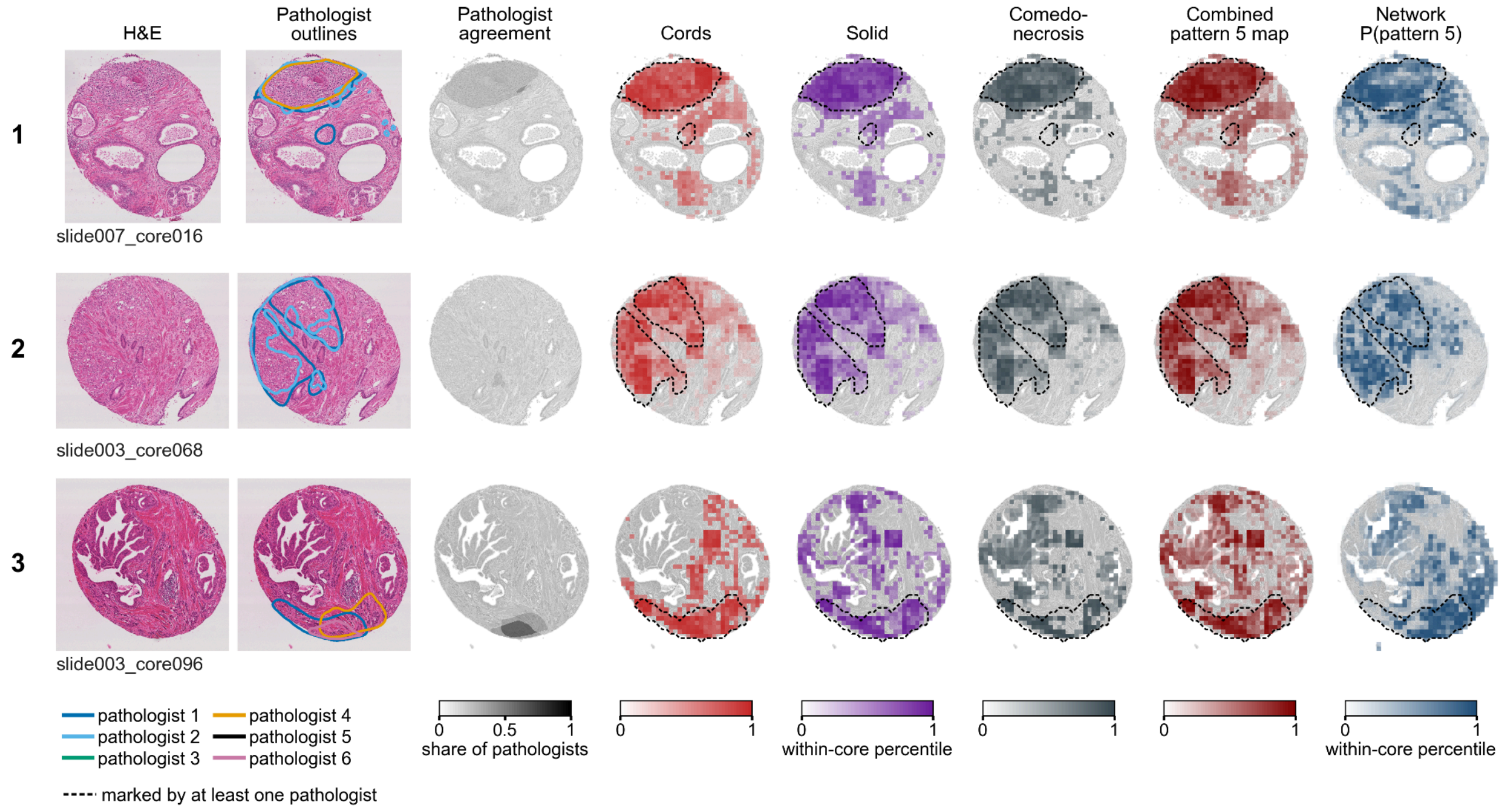

*Figure 5: Pattern 5 on Gleason2019 cores. Pathologist outlines beside the cord, solid-group and comedonecrosis maps, the pattern 5 map and the pattern 5 probability of the grading network.*

## 4.5 Effect of named features on grading

Table 9 summarizes the removal of named features before the grading network. On the PANDA-Radboud test slides, removing the pattern 3 set reduces pattern 3 calls by 97.6%, the pattern 4 set reduces pattern 4 calls by 48.7%, and the pattern 5 set reduces pattern 5 calls by 89.9%. On the random subset of 100 test slides, the same removals reduce the calls by 97.8%, 54.4% and 88.0%, while the 20 random sets of unnamed features change them by −9.5% to +11.1%, −8.3% to +11.9% and −14.9% to +31.1%, correspondingly. Each named removal ranks first among the 21 removals of its pattern (Figure 6).

Within pattern 4, removing the cribriform set alone takes away 32% of the calls and the poorly formed set 33%, and within pattern 5, the cord set takes away 69%, the solid-group set 81% and the comedonecrosis set 65%. Removing the well-formed set raises pattern 4 calls by 52% and pattern 5 calls by 19%, which suggests that tissue without its pattern 3 evidence is called a higher pattern. Half of the pattern 4 calls survive the removal of both named sets. Pattern 4 is the broadest of the three patterns, glomeruloid glands have no set of their own, and part of the network's pattern 4 evidence lies in unnamed features.

The 69 named features make up 3.4% of the code, yet they carry a median of 0.42 of the activation on tumor tokens of the PANDA-Radboud test slides, against 0.12 on other tissue (Table 3). The input share is similar, with medians of 0.371, 0.385 and 0.414 on tokens called pattern 3, 4 and 5 (Table 9). These shares describe how much of the input the named features make up, and the removal measures how much the grade depends on them.

*Table 9: Removal of named features.*

| Pattern | Named features | Named removal, 100 slides | Random range, 100 slides | Rank of 21 | Named removal, 506 slides | Input share |
|---|---|---|---|---|---|---|
| Pattern 3 | 16 | −97.8% | −9.5% to +11.1% | 1 | −97.6% | 0.371 |
| Pattern 4 | 34 | −54.4% | −8.3% to +11.9% | 1 | −48.7% | 0.385 |
| Pattern 5 | 20 | −88.0% | −14.9% to +31.1% | 1 | −89.9% | 0.414 |

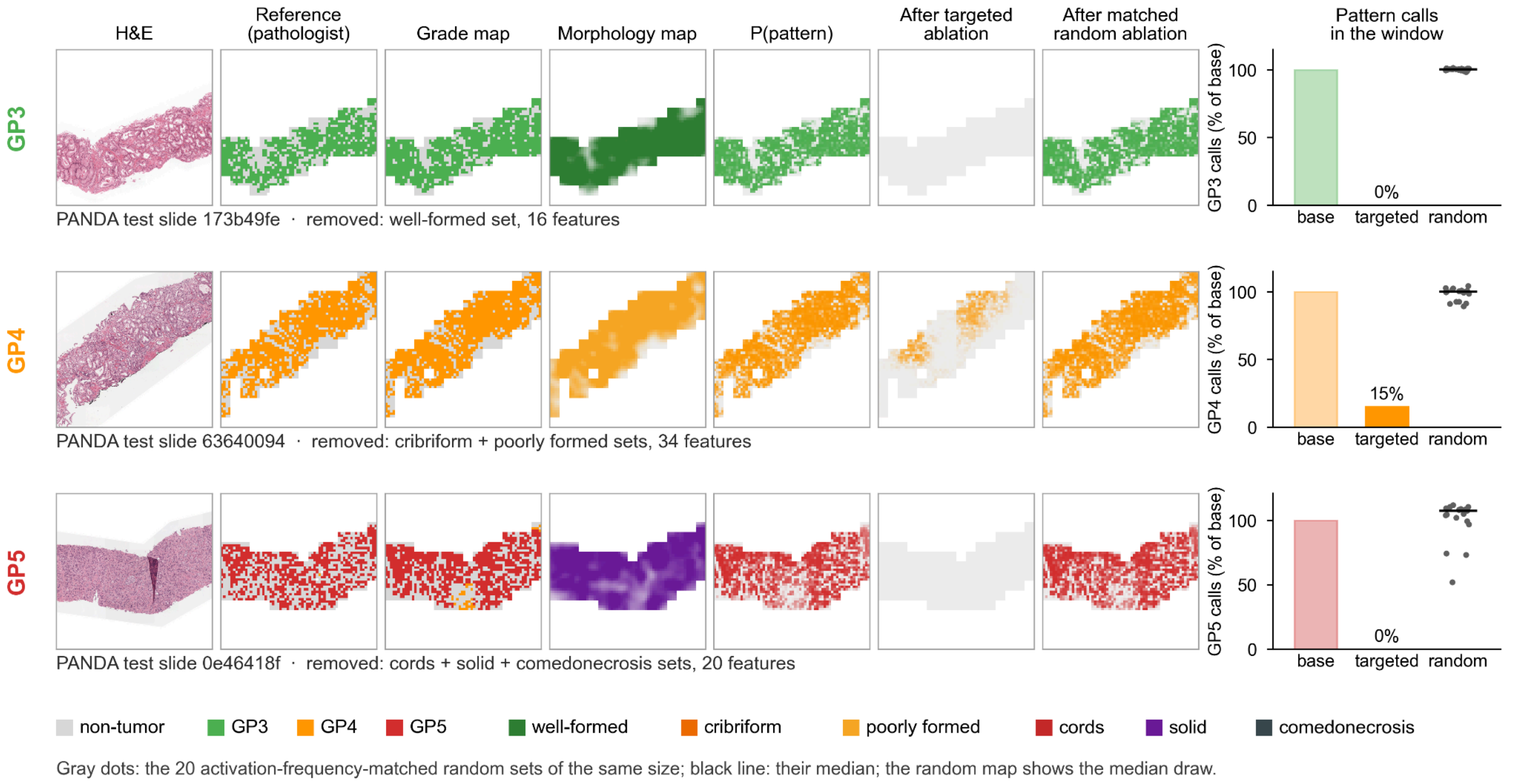

*Figure 6: Removal of named features. Named sets of a pattern are removed before the grading network and compared with random sets of unnamed features matched in size and activation frequency. Each example shows the morphology map, the pattern probability, and the probability after a named and after a random removal.*

# 5 Discussion

The removal experiments show that the grading network relies on the named features of the code when it assigns a pattern. Removing the named features of pattern 3 or pattern 5 takes away most of the corresponding calls, while none of the 20 random sets of unnamed features with the same size and activity come close. This is the main difference from post-hoc explanations, such as saliency maps or concept probes, which describe an already trained model from the outside, meaning that a highlighted region does not necessarily take part in the decision. Feature removal, on the other hand, acts directly on the input of the grading network, so any change in the calls can be attributed to the removed evidence. As for pattern 4, the fact that half of its calls survive indicates that part of its evidence lies in unnamed features, which is expected, since pattern 4 is the broadest pattern and glomeruloid glands have no set of their own.

Apart from the grading itself, the proposed code is transferable, stable and robust. Although it was pretrained only on biopsies from a single center, it keeps its sparsity and a median reconstruction error of 0.16 to 0.27 on TMA cores, on image patches from an external hospital and on three different scanners. Five of the six morphology maps also keep their accuracy across seven training runs, three of which start from independently pretrained SAEs. This is not obvious, since SAEs trained on the same data with different seeds are known to learn different features (Paulo and Belrose, 2025), and it indicates that the named morphologies follow the data and the labels, independently of the learned dictionary. The maps transfer without retraining to the biopsies of an external hospital and to unseen cores from another dataset, while on an independent cohort graded by three pathologists, the GG of the system agrees with them close to how they agree with each other, without any parameters tuned on that cohort.

Compared to Mittmann et al. (2025), where a segmentation network is trained on the explanation labels of pathologists, the morphology maps here are read from a general code without any trained segmenter, transferred to other datasets, and have their role in the grade measured directly on biopsies. Concept bottleneck models (Koh et al., 2020) route the prediction through the concepts alone, whereas our grading network reads the full code and still relies on its named part. Previous pathology studies with SAEs described tissue and morphology features (Le et al., 2024; Zhao et al., 2026), built slide classifiers on concepts named after discovery (Sun et al., 2026) and intervened on them (Patra et al., 2026). To the best of our knowledge, sparse features have not been aligned with multi-reader annotations of Gleason morphologies before, nor has the dependence of Gleason pattern predictions on named morphological features been measured against matched random controls. Regarding the transparency and human oversight that European guidance and the AI Act ask for, the proposed system offers the pathologist two checks on any slide, the morphology maps behind a grade and the change of the grade when their features are removed.

The morphology vocabulary of the system is bounded by the available annotations. Three of the nine annotated morphologies did not receive a set, since the public annotations contain at most 1362 tokens marked by two readers for each, and comedonecrosis, the smallest set with five features, is the least stable across runs. Given more multi-reader annotations, the same alignment can name these morphologies too, as nothing in the method is specific to the six used here. Similarly, the overlap of the solid-group and comedonecrosis sets with cords mirrors their co-occurrence in the readers' drawings, which is why pattern 5 is read from the combined map, and this map transfers to unseen cores. The unnamed part of the code is kept on purpose, so that the grading network does not lose evidence that the labels cannot name yet. Beyond the vocabulary, two limits bound the findings. Grading agreement falls on one of the three scanners, yet the code keeps its sparsity and reconstruction error on those scans and the embedding U-Net is affected in more comparisons, which suggests that the loss arises before the sparse code. In addition, the evaluation rests on one encoder and on retrospective public datasets, so other encoders and prospective use remain to be tested.

# 6 Conclusions

A single sparse code over FM tokens grades prostate biopsies with the competency of a dense representation, while the morphological evidence behind each grade can be read, checked against pathologists and tested by removal on any slide. The incorporation of larger multi-reader resources, which would give the rare morphologies and more of the pattern 4 evidence sets of their own, is a promising direction. Another interesting direction would be to include scans from more instruments during the pretraining of the SAE, so that grading becomes less sensitive to the scanner. Finally, a reader study, in which pathologists grade using the maps and the removal at hand, would measure the value of the system in clinical practice.

# Funding

This work was co-financed by the European Union and national funds through the National Strategic Reference Framework (NSRF) 2021–2027, under the Operational Programme "Competitiveness", Action "Bilateral Scientific and Technological Cooperation Greece–China" (Call Code 12KE, State Aid Code 11434), within the project "Research of Digital Pathology Image-based Intelligence Diagnosis Techniques of Prostate Cancer and Development of DT Software System - PRODIGY" (Grant Code: ΔΣΕΚΡ01-0021403, ERDF / European Regional Development Fund).

The funding sources had no involvement in the study design, in the collection, analysis and interpretation of data, in the writing of the report, or in the decision to submit the article for publication.

# Data availability

This study is a secondary analysis of publicly available, de-identified datasets used under their licenses. Approvals for these datasets are reported in their original publications.

The trained models are derived from UNI2-h embeddings, whose license (CC BY-NC-ND 4.0) restricts the distribution of derivatives, and are therefore not shared. The code is available from the corresponding author on reasonable request.